\documentclass[onecollarge]{svjour2}
\usepackage{graphicx}
\usepackage{mathptmx}      

\usepackage[T1]{fontenc}
\usepackage{bm}
\usepackage{amsmath}
\usepackage{cite}
\journalname{Few-Body Systems}

\begin{document}
\title{Experimentally accessible invariants encoded in interparticle correlations of harmonically trapped ultra-cold few-fermion mixtures}
\titlerunning{Invariants encoded in correlations of a few ultra-cold fermions}

\author{Daniel P{\k e}cak \and Mariusz Gajda \and Tomasz Sowi\'nski}

\institute{D. P{\k e}cak \and M. Gajda \and T. Sowi\'nski \at
              Institute of Physics, Polish Academy of Sciences,  Aleja Lotnikow 32/46, PL-02668 Warsaw, Poland \\
              \email{pecak@ifpan.edu.pl}
}

\date{Received: date / Accepted: date}

\maketitle

\begin{abstract}
A system of a two-flavour mixture of ultra-cold fermions confined in a one-dimensional harmonic trap is studied.
Using the well-known properties of the centre-of-mass frame we present a numerical method of obtaining energetic spectra in this frame for an arbitrary mass ratio of fermionic species.
We identify a specific invariant encoded in many-body correlations which may be helpful to determine an eigenstate of the Hamiltonian and to label excitations of the centre of mass.
The tool presented can be easily applied and thus may be particularly useful in an experimental analysis of the interparticle interactions which do not affect the centre of mass excitations in a harmonic potential.
\end{abstract}

\section{Introduction}
Few-body problems has been recently vastly explored theoretically \cite{Blume2010TwoThreeToMany,Blume2012FewBodyTraps,VolosnievNC2014}.
Partially it is due to the experiments on ultra-cold atoms making possible to verify theoretical concepts that had been infeasible before \cite{lewenstein2012ultracold}.
Recent developments in the field have shown that it is possible to control many parameters such as the number of atoms or the strength of mutual interactions with an exceptional precision \cite{wenz2013fewToMany,zurn2012fermionization,serwane2011tunableFermionSystem,MurmannPRL2015a,Kaufman2015Entangling,Onofrio2016Mixtures}.
Apart from adjusting these parameters, also the geometry of an external trap can be tuned.
In consequence, it is even possible to study effectively one-dimensional systems having very interesting and unique properties \cite{lewenstein2012ultracold}.

Because of their complexity, few-body systems are usually very challenging to treat analytically.
Also numerical studies may be very time-consuming due to a quite large number of degrees of freedom, particularly for fermionic species.
Also the straightforward mean-field methods do not work well for a small number of particles.
Therefore, in general, systems with mesoscopic number of particles are very difficult to treat theoretically.

One way to overcome these problems is to simplify the description by introducing variables directly related to the natural degrees of freedom of the system.
It is known that these generalized variables, in which description of the system is very simple, do exist.
However, even for small number of particles one needs to perform not obvious transformations
\cite{Busch1998, LiuHuDrummond2010ThreeAttractive, KestnerDuan2007ThreeBody, WernerCastin2006UnitaryThreeBody, loft2015variational, Koscik2012, Koscik2015vonNeuman, Koscik2016Bipartite, Olshanii20154Body, Harshman2016I, Harshman2016II}.
For example, for four particles it is convenient to introduce Jacobi coordinates \cite{Dehkharghani2016Impenetrable,Garcia-MarchPRA2014}.
The situation is even worse for larger number of particles and in practice these transformations are  very hard to implement.
Apart from these problems, in the case of quantum description, additional complexity appears because of the indistinguishability of particles and quantum uncertainty of the center of mass position.
If this uncertainty is relatively large, what can happen in attractive systems, then experimental separtion of relative dynamics from the center of mass motion becomes not trivial and has to be related to higher order correlations \cite{Gajda2006}.
On top of this, the symmetrization or anti-symmetrization of the wave function, when expressed in the language of generalized variables leads to the highly non-trivial constrains in the many-body picture.

Moreover, the generalized coordinates are not well suited for experiments.
Since atomic microscopes detect directly positions of particles.
Therefore, generalized variables are not helpful in the interpretation of experimental outputs.
It would be most convenient if a theoretical description of the system could be done in the most natural variables, i.e., positions of particles measured in experiments.
Unfortunately, in this case a lot of redundant information comes from trivial dynamics of the centre of mass of the system which is completely insensitive to interparticle interactions.
This general problem can be overcome in the case of the harmonic confinement, where the centre-of-mass degree of freedom can be separated out from the relative motion \cite{Busch1998,LiuHuDrummond2010ThreeAttractive, KestnerDuan2007ThreeBody, WernerCastin2006UnitaryThreeBody, Garcia-MarchPRA2014, loft2015variational,lawson1974, FedorovMikkelsen2015Recombination, Amin2016CorrelationsHO,Gajda2000COM,Bienias2014Rydberg, Brandt2016COMconvergence,Blakie2016PCS,Bradley2016Brownian}.
Here, we study one-dimensional fermionic mixtures and we show how to distillate the relevant information about relative excitations encoded in the full many-body wave functions.
Particularly, we identify experimentally accessible quantity which are invariant under the change of the centre of mass excitations.

Additionally, we present a simple numerical method of extracting the eigenstates of the centre of mass and the relative motion from particular eigenstate of the full many-body Hamiltonian.
In consequence, it is possible to simplify the description of the many-body spectrum.
The method is based on a trivial properties of the centre-of-mass motion, however the procedure may shed some light on the problem of few-body systems.
We also show a technique of obtaining information about the relative excitations from the simple two-body correlation, namely the two-body density profile.
We believe that it may be a useful tool in the field of experimental few-body physics where the two-body correlations are experimentally available to observe.
We show that the procedure is very general and can be used for fermionic mixtures of different masses.

The article is composed as following.
In the next Sec.~\ref{sec:model} we describe the system under consideration.
Then we introduce the frame of the centre of mass in Sec.~\ref{sec:COM} which is followed by the description of our numerical methods in Sec.~\ref{sec:method}.
The subsequent Sec.~\ref{sec:spectrum} contains the analysis of the spectra in the centre-of-mass frame.
Then in Sec.~\ref{sec:correlations} a method of decoding an information from the interparticle correlations is presented.
Finally, the last Sec.~\ref{sec:conclusions} concludes our paper.

\section{The model} \label{sec:model}
Let us consider a one-dimensional system of two distinguishable types of fermions confined in an external harmonic potential of a frequency~$\omega$.
We will label them with the quantum number $\sigma\in\{\downarrow,\uparrow\}$.
The quantum number $\sigma$ does not necessary mean that the particles differ with the spin projection.
In fact they may not have the spin degree of freedom at all.
We assume that $\sigma$ is conserved and it is not a dynamical variable.
Experimentally, this can be realized for example by choosing two proper hyperfine states of a nucleus \cite{wenz2013fewToMany} or two different elements \cite{Wille6Li40K,tiecke2010Feshbach6Li40K}.
We model interactions between particles of different flavours with a $\delta$-like contact potential \cite{PethickSmith} which is well justified in the ultra-cold regime and causes no mathematical problems in one dimension (there is no need of regularization).
In this approximation, the interaction within a given flavour vanishes due to the Pauli exclusion principle.
The Hamiltonian of the system reads:
\begin{equation} \label{eq:hamiltonian}
 \hat{H}  =  \sum_{i=1}^{N_{\downarrow}} \left[ - \frac{\hbar^2}{2 m_{\downarrow}} \frac{\partial^2}{\partial x_i^2} + \frac{m_{\downarrow} \omega^2}{2} x_i^2 \right]
 + \sum_{j=1}^{N_{\uparrow}} \left[ - \frac{\hbar^2}{2 m_{\uparrow}} \frac{\partial^2}{\partial y_j^2} + \frac{m_{\uparrow} \omega^2}{2} y_j^2 \right]
 + g_{\mathrm{1D}} \sum_{i=1}^{N_{\downarrow}}\sum_{j=1}^{N_{\uparrow}} \delta(x_i - y_j),
\end{equation}
where $N_{\downarrow}, N_{\uparrow}$ are the numbers of fermions with masses $m_{\downarrow}$ and $m_{\uparrow}$, respectively.
The Hamiltonian of the system ~(\ref{eq:hamiltonian}) commutes with the operators of particle numbers, $[\hat{H},\hat{N}_{\uparrow}]=[\hat{H},\hat{N}_{\downarrow}]=0$, what implies that the number of particles of each kind is conserved and fermions cannot change their flavour.
In real experiments this condition is met.
In the realizations using two different hyperfine states the interactions channels that change the quantum number $\sigma$ are practically disabled.
For a mass imbalanced system, there is an additional superselection rule resulting from the conservation of the mass that forbids changing atoms of one species to another.
An effective one-dimensional coupling parameter $g_{\mathrm{1D}}$ can be obtained from the full three-dimensional theory of scattering by integrating out two dimensions as shown in \cite{Olshanii1998}.
What is worth noticing, the coupling $g_{\mathrm{1D}}$ may be tuned experimentally, for example, by varying an external magnetic field \cite{Wille6Li40K,tiecke2010Feshbach6Li40K}.

We assume the same trapping frequency $\omega$ for both species which implies the same energy scale $\hbar\omega$ independently on their masses.
In the following, we choose to measure all the masses in the unit of $m_{\downarrow}$ and consequently the unit of length is $\sqrt{\hbar/(m_{\downarrow}\omega)}$.
The Hamiltonian (\ref{eq:hamiltonian}) in a dimensionless form reads:
\begin{equation} \label{eq:hamiltonianDimensionless}
 \hat{H}  =  \sum_{i=1}^{N_{\downarrow}} \left[ - \frac{1}{2 } \frac{\partial^2}{\partial x_i^2} + \frac{1}{2} x_i^2 \right]
 + \sum_{i=1}^{N_{\uparrow}}   \left[ - \frac{1}{\mu} \frac{1}{2 } \frac{\partial^2}{\partial y^2_i} + \mu \frac{1}{2} y^2_i \right] + g \sum_{i=1,j=1}^{N_{\downarrow},N_{\uparrow}} \delta(x_i - y_j),
\end{equation}
where we introduced the mass ratio of the particles, $\mu = m_{\uparrow}/m_{\downarrow}$, and the dimensionless interaction strength coupling, $g=g_{\mathrm{1D}} \sqrt{m_{\downarrow}/(\hbar^3 \omega)}$.

Before we apply our numerical techniques, it is extremely convenient to rewrite the Hamiltonian (\ref{eq:hamSecondQ}) in the second quantization form:
\begin{equation}\label{eq:hamSecondQ}
 \hat{H}  =  \sum_{\sigma} \int\mathrm{d}x \hat{\Psi}_{\sigma}^{\dag}(x) \hat{H}_{\sigma}(x) \hat{\Psi}_{\sigma}(x)
 + g \int\mathrm{d}x \hat{\Psi}_{\uparrow}^{\dag}(x) \hat{\Psi}_{\downarrow}^{\dag}(x) \hat{\Psi}_{\downarrow}(x) \hat{\Psi}_{\uparrow}(x),
\end{equation}
where appropriate single-particle Hamiltonians $\hat{H}_{\sigma}(x)$ are defined as:
\begin{subequations}
\begin{eqnarray}
H_{\downarrow}(x) &=& - \frac{1}{2 } \frac{\mathrm{d}^2}{\mathrm{d} x^2} + \frac{1}{2} x^2, \\
H_{\uparrow}(x)   &=& - \frac{1}{2 \mu} \frac{\mathrm{d}^2}{\mathrm{d} x^2} +  \frac{\mu}{2} x^2.
\end{eqnarray}
\end{subequations}
A field operator $\hat{\Psi}_{\sigma} (x)$ annihilates a particle of a type $\sigma$ in a position $x$.
Within the same flavour, for the indistinguishable fermions, the anticommutation relations hold $\{\hat{\Psi}_{\sigma} (x), \hat{\Psi}_{\sigma}^\dag (x')\} = \delta(x-x')$ and $\{\hat{\Psi}_{\sigma} (x), \hat{\Psi}_{\sigma} (x')\} = 0$.
For the particles of different flavours, the choice of commutation relation is not unique, since the physical results do not depend on this choice (as long as the relations are used consistently).

\section{The centre-of-mass frame} \label{sec:COM}
In the classical mechanics as well as in the quantum mechanics \cite{IBB1985COM,IBB2002COM,Sowinski2007Linear} there is a particular class of problems which are described by a Hamiltonians having biquadratic form of positions and momenta.
In this case, if the interaction between particles depends only on their relative positions, it is possible to canonically transform dynamical variables in such a way that the centre-of-mass motion separates from the relative motion of the system.
The motion of the centre of mass does not depend on the interaction which influences only the motion of the remaining degrees of freedom.
The two-body problem \cite{Busch1998} is an example but the separation can also be done for larger number of particles \cite{LiuHuDrummond2010ThreeAttractive, KestnerDuan2007ThreeBody, WernerCastin2006UnitaryThreeBody, FedorovMikkelsen2015Recombination, Amin2016CorrelationsHO} at the cost of using more sophisticated methods of describing the relative motion  \cite{loft2015variational,Dehkharghani2016Impenetrable}.
In practice, the mentioned transformation of variables is difficult to perform straightforwardly, however it is a standard tool in a theoretical analysis of few-body problems~\cite{Brandt2016COMconvergence,Blakie2016PCS,Bradley2016Brownian}.
In the following we will separate the the centre-of-mass motion within the scope of the second quantisation formalizm.

Position of the centre of mass for classical system of $N_\uparrow+N_\downarrow$ particles has a form:
\begin{equation} \label{eq:xcm}
 R_{\mathrm{CM}}  =  \frac{\sum_{i=1}^{N_{\downarrow}} x_i \cdot m_{\downarrow} + \sum_{j=1}^{N_{\uparrow}} y_j \cdot m_{\uparrow}}{N_{\downarrow} \cdot m_{\downarrow} + N_{\uparrow} m_{\uparrow}}
  =  (N_{\downarrow} + \mu N_{\uparrow})^{-1}
      \left(
      \sum_{i=1}^{N_{\downarrow}} x_i  + \sum_{j=1}^{N_{\uparrow}} \mu \cdot y_j
      \right).
\end{equation}
Since the quantity (\ref{eq:xcm}) is a linear function of the positions of particular particles, the corresponding quantum-mechanical counterpart, written in the second quantization picture, is defined as:
\begin{equation} \label{eq:rcm}
\hat{R}_{\mathrm{CM}}  =  (N_{\downarrow} + \mu N_{\uparrow})^{-1}
 \int\mathrm{d}x
    \left( \hat{\Psi}_{\downarrow}^{\dag} (x)  x \hat{\Psi}_{\downarrow}   (x)
	  + \mu \hat{\Psi}_{\uparrow}^{\dag} (x) x  \hat{\Psi}_{\uparrow} (x)
    \right).
\end{equation}
The field operators can be decomposed into the basis of the eigenfunctions of an appropriate harmonic oscillator $\phi_{i\sigma}(x)$:
\begin{equation}\label{eq:fieldOp}
 \hat{\Psi}_{\sigma}(x) = \sum_i\phi_{i\sigma}(x)\hat{a}_{i\sigma},
\end{equation}
where the operator $\hat{a}_{i\sigma}$ annihilates a fermion of type $\sigma$ from the orbital $i$ and the following condition is met $(H_{\sigma}(x)-\hbar \omega (i + 1/2)) \phi_{i\sigma}(x) = 0 $.
The sum (\ref{eq:fieldOp}) runs from zero to infinity.
For convenience we introduce the coefficients:
\begin{equation}
 X_{ij,\sigma} = \mu_{\sigma} \int  \mathrm{d}x \phi_{i \sigma}^{*} (x) x \phi_{j \sigma} (x) ,
\end{equation}
where $\mu_{\downarrow} = 1$ and $\mu_{\uparrow} = \mu$.
Then, the centre-of-mass operator gets simplified form:
\begin{equation}\label{eq:Rcm}
\hat{R}_{\mathrm{CM}}
 = (N_{\downarrow} + \mu N_{\uparrow})^{-1} \sum_{ij}\sum_{\sigma}
     X_{ij,\sigma} \hat{a}_{i \sigma}^{\dag} \hat{a}_{j \sigma}.
\end{equation}
Analogously, one introduces the operator of the total momentum:
\begin{equation}
\hat{P}_{\mathrm{CM}}
 = -i \int \mathrm{d} x
    \left( \hat{\Psi}_{\downarrow}^{\dag} (x)  \partial_x  \hat{\Psi}_{\downarrow} (x)
	  +  \hat{\Psi}_{\uparrow}^{\dag} (x)  \partial_x  \hat{\Psi}_{\uparrow} (x)
    \right),
\end{equation}
which can be rewritten in a decomposed form as:
\begin{equation}\label{eq:Pcm}
  \hat{P}_{\mathrm{CM}}
  = \sum_{ij} \sum_{\sigma}
  P_{ij,\sigma} \hat{a}_{i \sigma}^{\dag} \hat{a}_{j \sigma},
\end{equation}
where appropriate coefficients are defined as:
\begin{equation}
 P_{lm,\sigma}  =  - i \int  \mathrm{d}x \phi_{l \sigma}^{*} (x) \partial_x \phi_{m \sigma} (x).
\end{equation}
It can be shown that the operator of the total momentum $\hat{P}_{\mathrm{CM}}$ and the operator of the position of the centre of mass $\hat{R}_{\mathrm{CM}}$ are canonically conjugate variables and natural commutation relations between them hold, $[\hat{R}_{\mathrm{CM}},\hat{P}_{\mathrm{CM}}]=i\hbar$ and $[\hat{R}_{\mathrm{CM}},\hat{R}_{\mathrm{CM}}]=[\hat{P}_{\mathrm{CM}},\hat{P}_{\mathrm{CM}}]=0$.
The Hamiltonian of the centre of mass written in a dimensionless units reads:
\begin{equation} \label{eq:hamCM}
 \hat{H}_{\mathrm{CM}}
 = \frac{\hat{P}_{\mathrm{CM}}^2}{2(N_{\downarrow} + \mu N_{\uparrow})}
 + \frac{(N_{\downarrow} + \mu N_{\uparrow})}{2} \hat{R}_{\mathrm{CM}}^2 .
\end{equation}
As suspected, the Hamiltonian (\ref{eq:hamCM}) describes a particle of the mass equal to the total mass of the system $(N_{\downarrow} + \mu N_{\uparrow})$ trapped in the harmonic potential.
As a consequence, the total Hamiltonian can be separated to $\hat{H}  = \hat{H}_{\mathrm{CM}} + \hat{H}_{\mathrm{REL}}$.
$\hat{H}_{\mathrm{REL}}$ represents the relative Hamiltonian and only this part contains all information about internal interactions between ultra-cold fermions.
It is also easy to check that the total Hamiltonian of the system $\hat{H}$ and the Hamiltonian of the centre of mass $\hat{H}_{\mathrm{CM}}$ as well as the interaction Hamiltonian $\hat{H}_{\mathrm{REL}}$ do commute, i.e., $[\hat{H},\hat{H}_{\mathrm{CM}}] = [\hat{H},\hat{H}_{\mathrm{REL}}] = 0$.
It means that there exists a basis in the Hilbert space in which all three Hamiltonians have simultaneously diagonal form.

Every eigenstate of the Hamiltonian (\ref{eq:hamSecondQ}) can be denoted by the set of $N=N_{\downarrow}+N_{\uparrow}$ quantum numbers.
Since the centre of mass is separated out, the remaining $N-1$ quantum numbers describe the relative degrees of freedom of the system.
Therefore, the total energy $E$ of a given eigenstate of the Hamiltonian (\ref{eq:hamSecondQ}) can be written in a form:
\begin{equation}
 E(\nu_1,\nu_2,\ldots,\nu_N) = \underbrace{E(\nu_1)}_{(\nu_1 + \frac{1}{2})} + E(\nu_2,\nu_3,\ldots,\nu_{N}),
\end{equation}
where ${\bm \nu} = (\nu_1;\nu_2,\ldots,\nu_N)$ is a set of quantum numbers describing a given eigenstate of the Hamiltonian $\hat{H}$ and $E(\nu_1)$ corresponds to the centre-of-mass energy present in this state.
Naturally, for a given set of quantum numbers $(\nu_2,\nu_3,\ldots,\nu_{N})$ describing the relative motion excitations, there are infinitely many states with different values of $\nu_1$ with energies separated by the excitations of the centre of mass.

The method of getting the energy spectrum of the Hamiltonian (\ref{eq:hamSecondQ}) filtered out from the centre-of-mass excitations is based on a quite trivial observation that the eigenstates of the Hamiltonian $\hat{H}$ are composed as products of eigenstates of $\hat{H}_{\mathrm{CM}}$ and $\hat{H}_{\mathrm{REL}}$.
Therefore, we can find the eigenvalue $E(\nu_1)$ of any eigenstate of $\hat{H}$ by calculating an expectation value of the Hamiltonian of the centre of mass $\hat{H}_{\mathrm{CM}}$ in this state.
Then, by collecting all eigenstates of the total Hamiltonian $\hat{H}$ with the same eigenvalue $E(\nu_1)$, we get the spectrum that contains only one specific excitation of the centre of mass.
The choice of the quantum number $\nu_1$ is arbitrary, but for technical reasons (our method has the best accuracy in the low energy regime) the best choice is to gather all eigenvalues for $\nu_1 = 0$ corresponding to the ground state of the centre of mass with the eigenenergy $\hbar\omega/2$.

\section{Method} \label{sec:method}
\begin{figure}
  \begin{center}
\resizebox{\textwidth}{!}{%
  \includegraphics{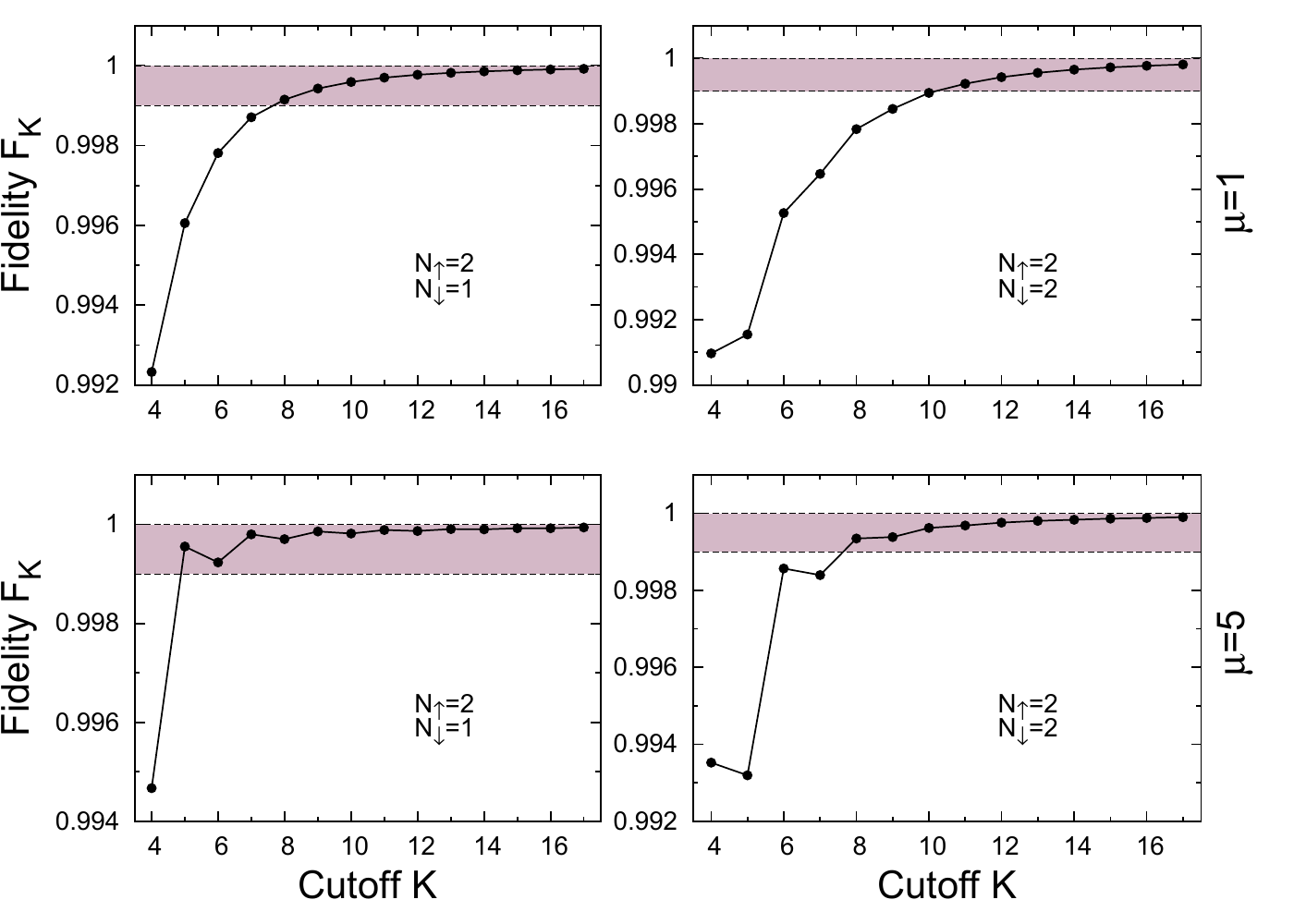}
}
\end{center}
\caption{The fidelity $\mathbf{F}_K$ calculated for the many-body ground-state of the system for different number of particles and different mass ratios $\mu$ as a function of the cutoff $K$. Shaded area marks the region where the fidelity deviates from one by less than $10^{-3}$. All results are presented for $g=3$.
\label{fig:errors}}
\end{figure}

The Hamiltonian (\ref{eq:hamSecondQ}) does not couple the Hilbert subspaces of different particle numbers, thus it can be diagonalized in each of those subspaces independently. The numerical method of the exact diagonalization of the Hamiltonian $\hat{H}$ returns the eigenenergies $E_i$ as well as the many-body eigenstates $|i\rangle$. Of course to apply the method, one has to introduce in the sum (\ref{eq:fieldOp}) a cutoff $K$ at sufficiently large index. It is justifiable because for any finite interaction $g$, only the lowest excitations play a role. In principle, we assume that the cutoff is appropriate if its further increase does not affect the results. In practice, it is done by calculating consecutive fidelities $\mathbf{F}_K(|i\rangle)$ defined for chosen many-body eigenstate $|i\rangle$ as the overlap between states $|i\rangle_K$ and $|i\rangle_{K+1}$, {\it i.e.}, the eigenstate calculated $|i\rangle$ for cutoffs $K$ and $K+1$. We accept the cutoff as a sufficiently large if the fidelity deviates from one by less than $10^{-3}$. In Fig. \ref{fig:errors} we show example adaptation of this procedure to the many-body ground-state of the system for different number of particles, different mass ratios $\mu$ and interaction $g=3$.

Since the two Hamiltonians $\hat{H}$ and $\hat{H}_{\mathrm{CM}}$ do commute, one can calculate the energy in the centre-of-mass frame using the Hamiltonian of the centre of mass $\hat{H}_{\mathrm{CM}}$ for any eigenstate of the Hamiltonian $\hat{H}$. Then by collecting all the states in the ground state of the centre-of-mass motion, i.~e. $\hat{H}_{\mathrm{CM}} |i\rangle = \hbar\omega/2|i\rangle$, one finally gets all states with excited internal motion only.

The situation is slightly more complicated if the eigenstates are $j$-fold degenerated.
Then, the whole set of the eigenstates $\{|m_1\rangle,\ldots,|m_j\rangle\}$ of the full Hamiltonian $\hat{H}$  has the same energy.
In these cases we calculate all matrix elements of the centre-of-mass Hamiltonian $\hat{H}_{\mathrm{CM}}$ in the basis of the degenerated eigenstates $\langle i| \hat{H}_{\mathrm{CM}}|i'\rangle$, where $|i\rangle \in\{|m_1\rangle,\ldots,|m_j\rangle\}$, and diagonalize it.
The lowest eigenstate we got after diagonalization is obviously an eigenstate of $\hat{H}_{\mathrm{CM}}$ and it is a superposition of some eigenstates of $\hat{H}$. This eigenstate has the lowest energy of the centre-of-mass excitation.
Therefore, this combination is the state we are interesting in.

Note, that in the procedure described above any assumptions about interactions, mass ratios or any additional symmetries are not made.
Thus the technique is very general.
In the following, the described method of filtering out the spectrum of the higher excitations of the centre of mass, is applied to the systems of equal masses $\mu=1$ as well as to the systems with mass imbalance $\mu\neq1$, bearing in mind that nowadays experiments on mixtures of different atomic species are also possible \cite{Wille6Li40K, tiecke2010Feshbach6Li40K}.

\section{Spectrum of the Hamiltonian}\label{sec:spectrum}
\begin{figure}
  \begin{center}
\resizebox{\textwidth}{!}{%
  \includegraphics{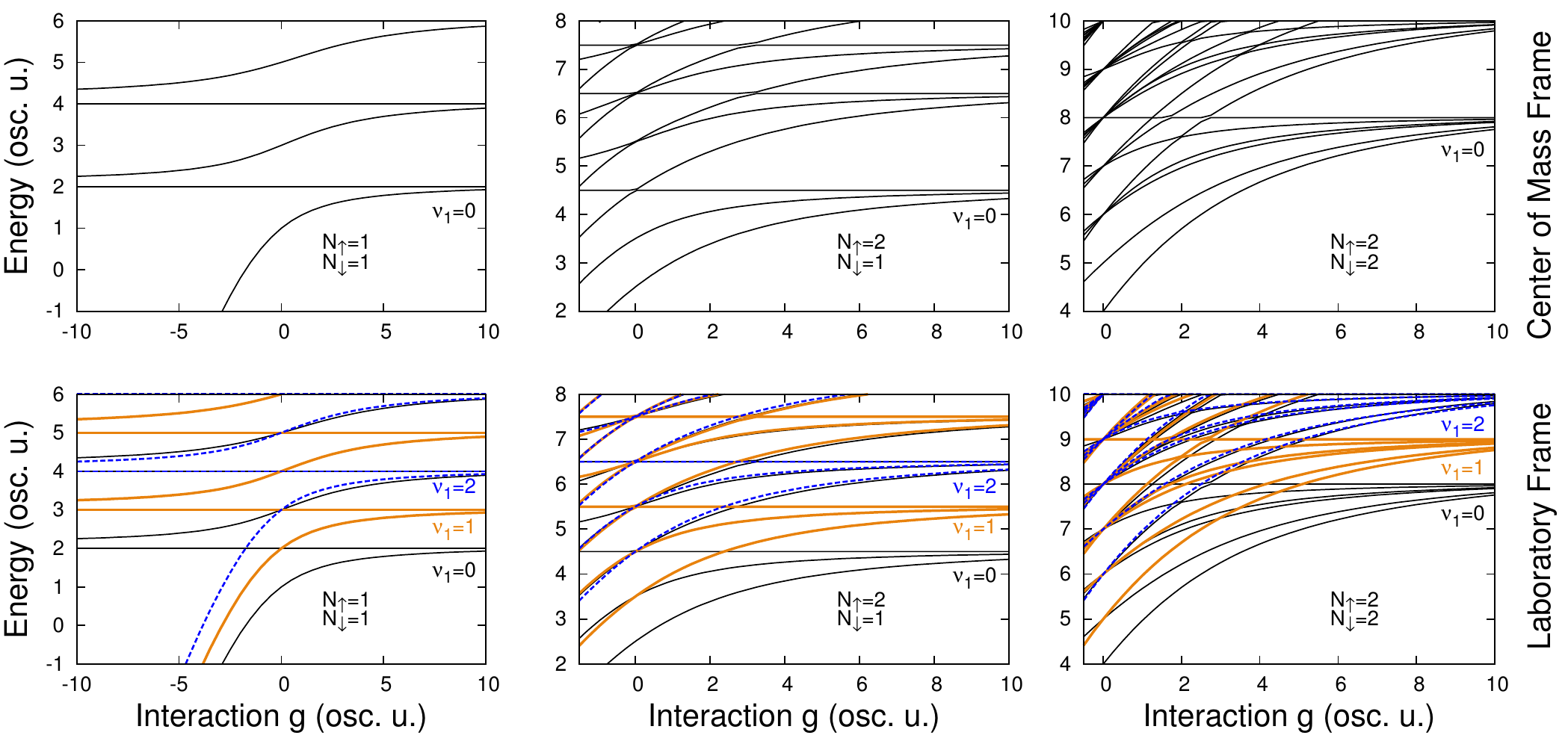}
}
\end{center}
\caption{The spectra of the same mass ($\mu=1$) fermions as functions of the interaction strength $g$ for different numbers of particles, $N_{\uparrow},N_{\downarrow}$.
The top panels correspond to the centre-of-mass frame spectrum.
They are used to construct corresponding full spectra presented in the bottom panels.
The black thin line corresponds to the ground state of the centre of mass ($\nu_1=0$), the thick orange line to the first excited state ($\nu_1=1$) and the blue dotted line to the second excited state of the centre of mass ground state ($\nu_1=2$).
\label{fig:spectrumM1}}
\end{figure}
First, we apply the procedure described in the previous section to the two-flavoured systems of several particles of the same mass $\mu=1$.
The results are shown in Figure~\ref{fig:spectrumM1}.
The top panels show the spectrum in the centre-of-mass frame (the top left panel reproduces the result obtained analytically in \cite{Busch1998}).
The eigenstates in the top panels are chosen to be in the ground state of the centre of mass, thus by shifting this plot by $\hbar\omega,2\hbar\omega,\ldots$ one can obtain the full spectrum of the system (shown on the bottom panels of Figure~\ref{fig:spectrumM1}) reproducing results well-known for the equal mass mixtures of fermions \cite{SowinskiGrass2013FewInteracting,Gharashi2013UpperBranchCorrelations}.
By increasing the number of particles, the spectrum becomes more complicated.
Even after simplification to the centre-of-mass frame, the spectrum is complex due to increasing number of degrees of freedom in a relative motion (right top panel of Figure~\ref{fig:spectrumM1}).

\begin{figure}
\resizebox{\textwidth}{!}{%
  \includegraphics{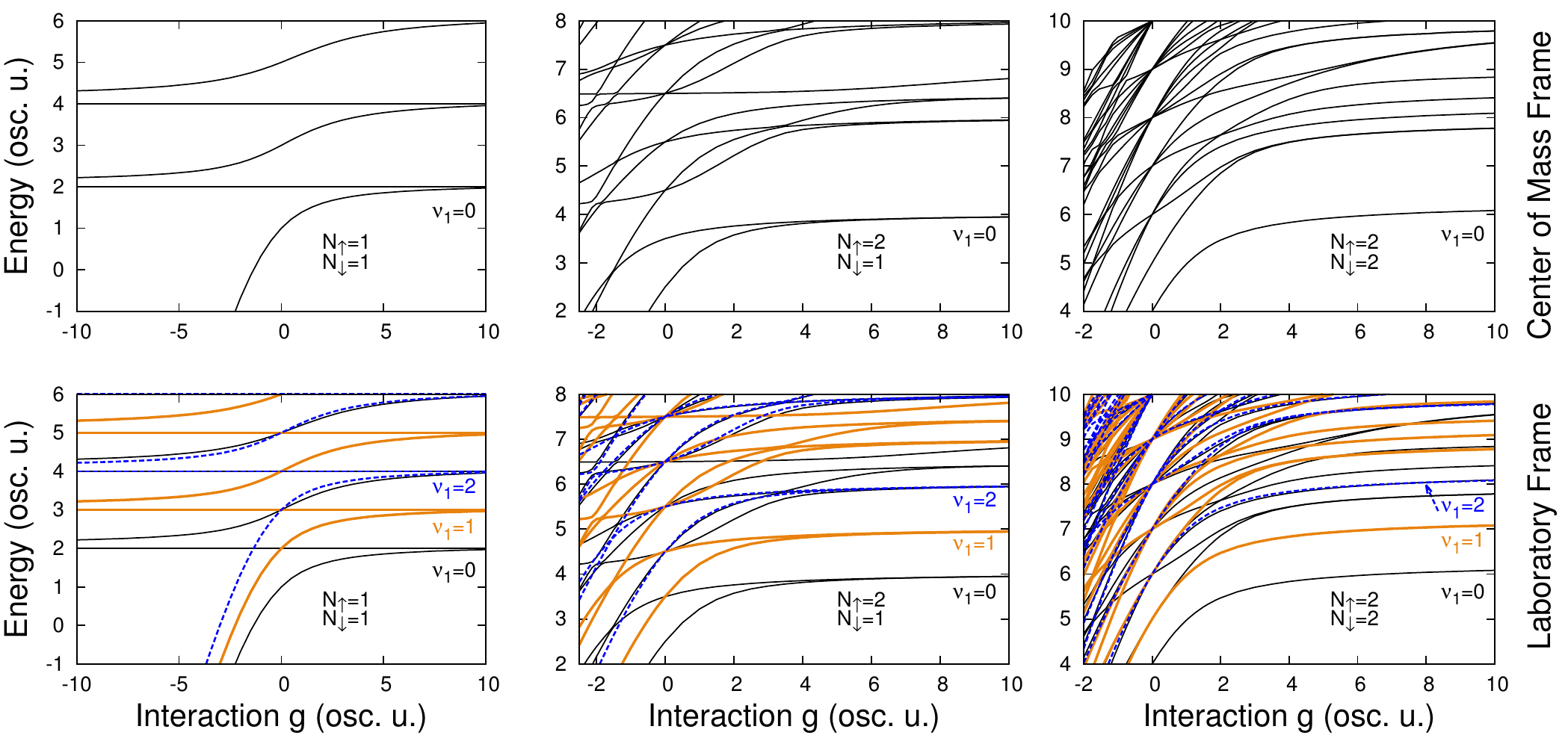}
}
\caption{The spectra of fermions with different mass ($\mu\neq1$) as functions of the interaction strength $g$ for different numbers of particles, $N_{\uparrow},N_{\downarrow}$.
The top panels correspond to the centre-of-mass frame spectrum.
They are used to construct full spectra presented in the bottom panels.
The black thin line corresponds to the ground state of the centre of mass ($\nu_1=0$), the thick orange line to the first excited state ($\nu_1=1$) and the blue dotted line to the second excited state of the centre of mass ground state ($\nu_1=2$).
\label{fig:spectrumM5}}
\end{figure}

Similar analysis can be done also for the mass-imbalanced mixtures as shown in Figure~\ref{fig:spectrumM5}.
The top panels show the spectrum in the centre-of-mass frame.
Shifting this spectrum by $\hbar\omega,2\hbar\omega,\ldots$ leads directly to the full many-body spectrum (bottom panels).
Apart from the trivial two-body case, the spectra in the centre-of-mass frame differ qualitatively from spectra of the equal mass systems.
For example, totally antisymmetric states do not exist in the mass-imbalanced case.
This is a direct consequence of different shapes of the single-particle orbitals (for more details see \cite{Pecak2016Separation,Pecak2016Transition}).

It is worth noticing that the two-particle problem is basically the same for equal mass and mass imbalanced cases.
In this particular case $N_{\uparrow}=N_{\downarrow}=1$, changing the mass of one component leads only to an effective change in the interaction strength.
It is quite obvious when one introduces the centre-of-mass position, $R_{\mathrm{CM}}=(x_1+\mu x_2)/(1+\mu)$, and the relative position, $r=(x_1-x_2)\sqrt{\mu/(1+\mu)}$.
Then, the centre-of-mass Hamiltonian $\hat{H}_{\mathrm{CM}}$ gets a form of a Hamiltonian of the standard harmonic oscillator.
Whereas, the relative Hamiltonian $\hat{H}_{\mathrm{REL}}$, due to a specific normalization of the relative position, has the form:
\begin{equation}\label{eq:hamInt2}
\hat{H}_{\mathrm{REL}}^{N=2} = -\frac{1}{2}\frac{\partial^2}{\partial r^2} + \frac{1}{2} r^2 + \sqrt{\frac{\mu}{1+\mu}}g \delta(r).
\end{equation}
It means that indeed, for any $\mu$ one can rescale the interaction strength, $ g \rightarrow g \sqrt{(1+\mu)/\mu}$, to map the different mass spectrum to the equal mass spectrum (compare top panels of Figure~\ref{fig:spectrumM1} and Figure~\ref{fig:spectrumM5}).
Note however, that although the energies are rescaled properly, other quantities, like for example single-particle densities, are different \cite{Pecak2016Separation}.

\section{Two-body correlations}\label{sec:correlations}
All the information about the state of the system is encoded in its wave function.
However, this mathematical object is very complex, especially for many-body states, due to a large number of degrees of freedom.
To get more insight into properties of a given quantum state, other simpler quantities, such as energy, single-particle density profile or correlations \cite{Manz2010G2}, may be considered to describe the properties of the system.
However, recent experiments allow one not only to take a 'photograph' of all ultra-cold atoms but also to measure all their positions simultaneously \cite{bakr2010probing,sherson2010single,OmranBloch2015PauliBlocking,Cheuk2015FermionicMicroscope,Cheuk2016MottMicroscope,Cheuk2016Correlations,ParsonsPRL2015,EdgePRA2015ImagingFermions,haller2015single,ParsonsPRL2015}.
It means that, in principle, it is possible to measure many-body correlations \cite{Gajda2016Imaging}.
This new tool gives another opportunity to probe theoretically predicted correlations in quantum systems and motivates us to describe closely the properties of higher correlations in few-body systems.

First, let us focus mainly on pair correlations.
Particularly, we analyze the properties of the system encoded in the pair-density profile, i.e. the diagonal part of the reduced two-body density matrix.
For a state $|i\rangle$ it is defined as:
\begin{equation}\label{eq:correlation}
 C(x,y) = \langle i| \hat{\Psi}_\downarrow^\dag(x) \hat{\Psi}_\uparrow^\dag(y) \hat{\Psi}_\uparrow(y) \hat{\Psi}_\downarrow(x) |i \rangle.
\end{equation}
This quantity is the simplest one that contains information about both components.
Integrating out any of the two variables would reduce the information to just one component, i.e., to the single-particle density profile.

\begin{figure}
\begin{center}
\resizebox{0.78\textwidth}{!}{%
  \includegraphics{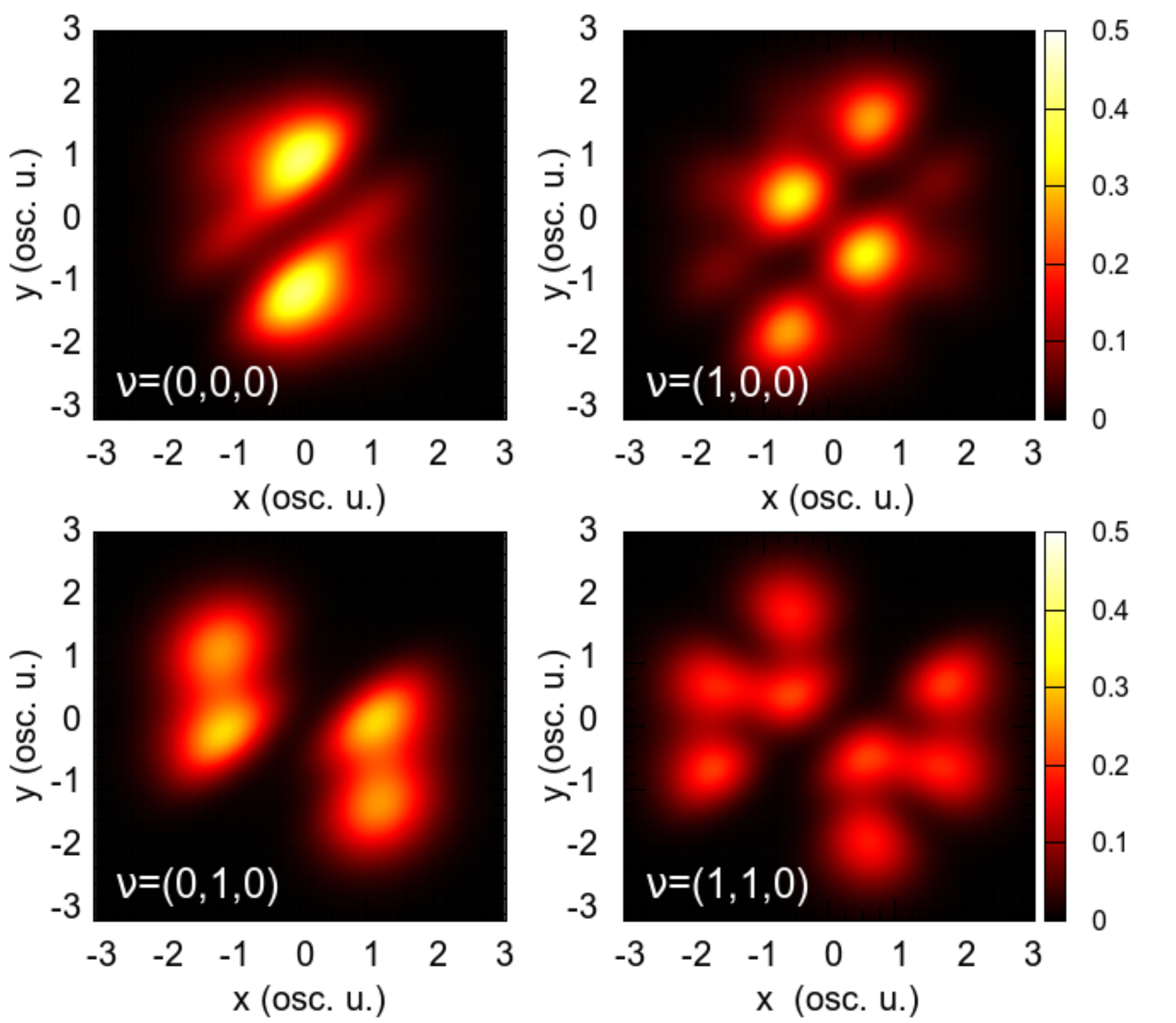}
}
\end{center}
\caption{Pair-density profiles of a few lowest eigenstates of a system $N_{\uparrow}=2,N_{\downarrow}=1$ for equal mass $\mu=1$ case and interaction $g=3$.
The left panel corresponds to the ground state of the centre of mass ($\nu_1=0$), while the right panel shows the first excited state of the centre of mass ($\nu_1=1$).
The top and the bottom rows show the ground state ($\nu_{2}=0$) and the first excited state ($\nu_{2}=1$) in the relative coordinates, respectively.
\label{fig:corrM1}}
\end{figure}

In Figure~\ref{fig:corrM1} we show the pair-density profile calculated for the ground state and one of the excited states of the system with $N_\uparrow=2,N_\downarrow=1$, and $g=3$ (upper left and bottom left panels).
The excited state is excited only in the internal motion.
In the right panels we show the same quantity calculated for the corresponding states with a single excitation of the centre of mass.
It is seen that based on the shape of the pair-density profile shown in Figure~\ref{fig:corrM1}, it is very hard to predict the set of quantum numbers ${\bm \nu}$.
It means that if one obtains such correlations experimentally for some states, it may be difficult to distinguish and properly classify them, especially if they contain unknown centre of mass or the relative motion excitations.

\begin{figure}
\begin{center}
  \resizebox{0.78\textwidth}{!}{%
  \includegraphics{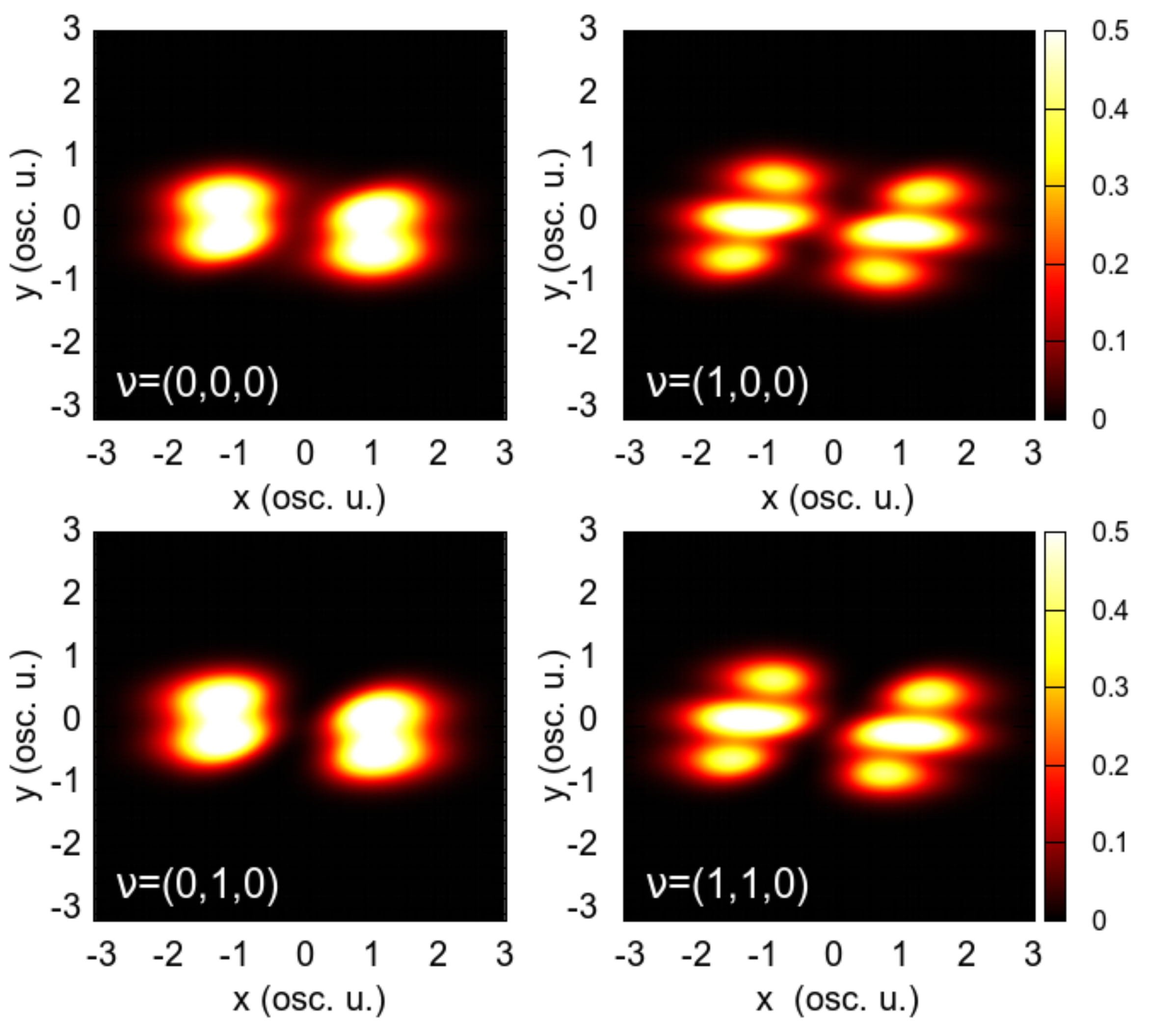}
  }
\end{center}
\caption{Pair-density profiles of a few lowest eigenstates of a system $N_{\uparrow}=2,N_{\downarrow}=1$ for mass $\mu=5$ case and interaction $g=3$.
The left panel corresponds to the ground state of the centre of mass ($\nu_1=0$), while the right panel the first excited state of the centre of mass ($\nu_1=1$).
The top and the bottom rows show the ground state ($\nu_{2}=0$) and the first excited state ($\nu_{2}=1$) in the relative coordinates, respectively.
\label{fig:corrM5}}
\end{figure}

The classification becomes even more confusing for the mass-imbalanced case.
The pair-density profiles for $\mu=5$ are shown in Figure~\ref{fig:corrM5}.
Here, in contrast to the equal mass case, states being in the same state of the centre-of-mass motion and differing only by relative excitations have almost the same profile of the pair density (left panels correspond to the ground state and right panels to the first excited state of the centre-of-mass motion).
The relative motion excitations change marginally the pair-density profile.
As a consequence, these states are in practice indistinguishable.
Additionally, the states (as seen in Figure~\ref{fig:spectrumM5}) are almost degenerated and have similar single-particle density profiles.
The only quantity that distinguishes these eigenstates is the centre-of-mass motion excitation.
However, much more information is encoded in the pair-density profiles.
As shown in the following, this hidden information on the centre of mass and the relative motion excitations can be extracted directly from the experimentally measured pair-density profiles.

The extraction is possible because one can separate some degrees of freedom in the system.
The separation of the centre of mass from the internal degrees of freedom means that a wave function of any eigenstate of the system can be presented as a product:
\begin{equation}\label{eq:WF_factor}
 \Psi(x_1,\dots,x_{N_{\downarrow}},y_1,\dots,y_{N_{\uparrow}}) = \Psi(R) \Psi(r_1,r_2,\dots,r_{N-1}),
\end{equation}
where $R$ denotes the centre of mass coordinate and $r_1,r_2,\dots,r_{N-1}$ are some general relative coordinates.
From all possible combinations of coordinate differences of the form $x_i-y_j,x_i-x_j$ and $y_i-y_j$, only $N_\uparrow+N_\downarrow-1$ are independent.
This implies that one has a lot of freedom in choosing relative distances between particles.
The relative distance $r_1 = x_1-y_1$ between a pair of fermions of different species is extremely useful and obviously, due to indistinguishability, does not depend on the choice of particular coordinate $x_i$ or $y_j$.

To explore the information hidden in two-body correlations of a state $|i\rangle$, let us consider the second moment of the pair-density profiles:
\begin{equation}\label{eq:Ir1}
{\cal I} = \int C(x,y) (x-y)^2 \mathrm{d}x \mathrm{d}y.
\end{equation}
It can be shown straightforwardly that this quantity does not depend on the centre of mass position $R$, i.e., it is insensitive to the excitations of the centre of mass of the system.
Therefore, as long as the eigenstates of the Hamiltonian (\ref{eq:hamSecondQ}) have the same value of relative degrees quantum numbers $(\nu_{2},\nu_{3}\ldots \nu_{N})$, they share the same value of ${\cal I}$.
Since ${\cal I}$ is invariant under changes of the centre of mass excitation, it is possible to divide all eigenstates of the system with respect to excitations of the relative motion.
It can be very useful when one is interested in preparing the state with a specific excitation of the internal motion but simultaneously is not able to control excitations of the centre of mass.
By calculating the invariant ${\cal I}$ from the experimentally accessible pair-density profile an internal excitation can be determined.
The method is quite useful since, as it is shown in Figure~\ref{fig:corrM1} and Figure~\ref{fig:corrM5}, the pair-density profile may have completely different shapes for states with different centre of mass excitations.
However, ${\cal I}$ extracted from these shapes is exactly the same.

\begin{figure}
  \resizebox{\textwidth}{!}{%
  \includegraphics{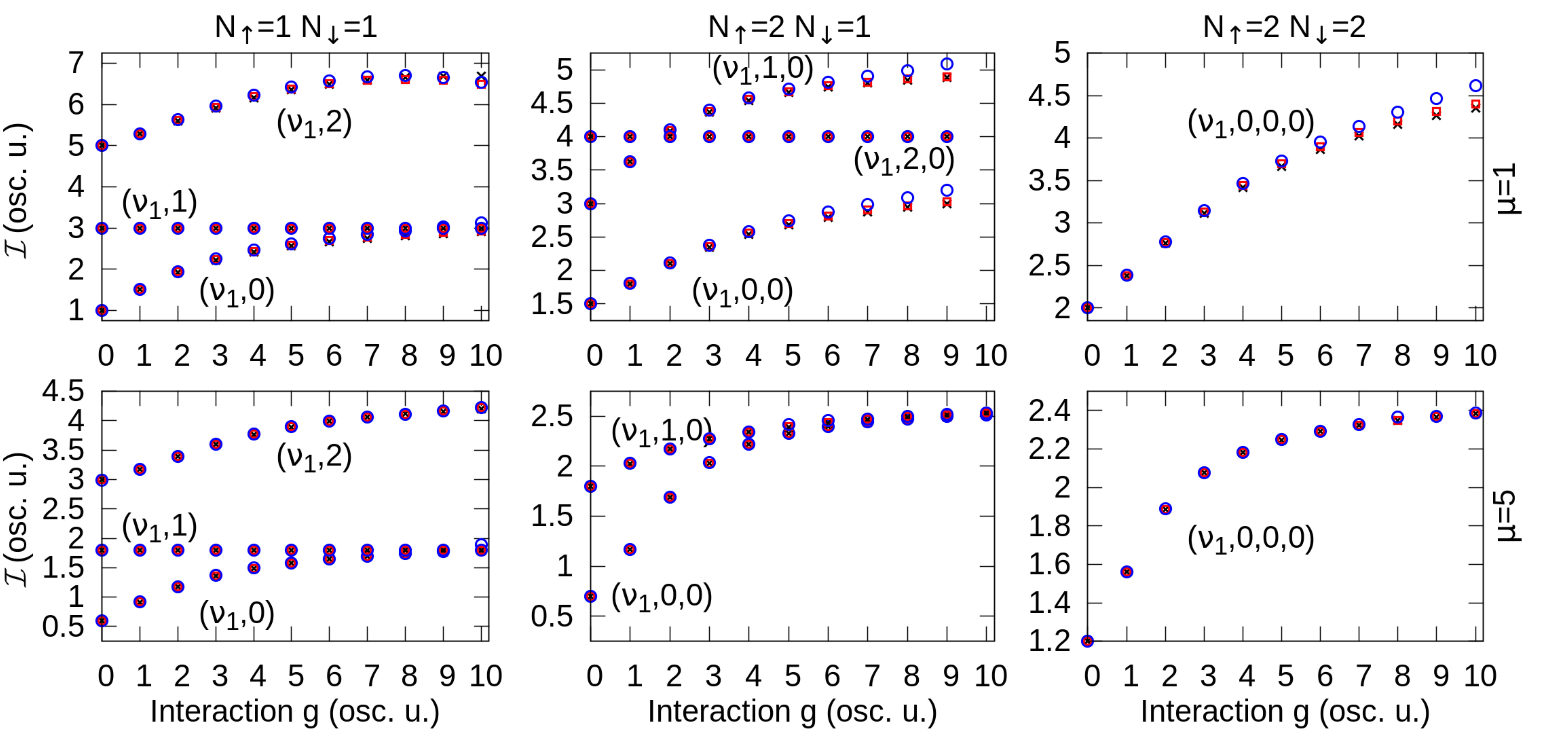}
  }
  \caption{
  The invariant ${\cal I}$ as a function of interaction strength $g$ for different eigenstates labeled by different quantum numbers $(\nu_1;\nu_2,\ldots,\nu_N)$.
  Top panels present the results for equal mass case ($\mu=1$), while the bottom panels for the mass imbalanced case ($\mu=5$).
  Consecutive columns shows results for different number of particles ($N_\uparrow+N_\downarrow=2,3,4$).
  Black crosses correspond to the states with the centre of mass in the ground state ($\nu_1=0$), red squares to the first excited state of the centre of mass ($\nu_1=1$), and blue circles to the second excited state ($\nu_1=2$).\label{fig:Ir1}}
\end{figure}
We calculated ${\cal I}$ for many different parameters (mass ratio $\mu$, number of particles $N_\uparrow, N_\downarrow$ and the interaction strength $g$).
The results are shown in Figure~\ref{fig:Ir1}.
The consecutive eigenstates of the centre of mass excitation $\nu_1=0,1$ and $2$ are marked by black crosses, red squares, and blue circles, respectively.
We found, as expected, that for a given set of quantum numbers describing the relative motion ($\nu_{2},\nu_3,\ldots,\nu_{N}$), the quantity ${\cal I}$ does not depend on the centre of mass excitations --- for a given interaction $g$ cross, square and circle are centred at the same point.
Some discrepancy may be seen for strong interactions $g$, especially for the equal mass case $\mu=1$ (top panels of Figure~\ref{fig:Ir1}).
This is an artefact of our numerical method caused by a finite size of the basis.
This deviation may be reduced by introducing larger single-particle basis (larger cutoff parameter), increasing simultaneously a time-cost of calculations.
Since the basis functions for mass-imbalanced systems are different for different species, therefore a numerical error for $\mu=5$ is smaller than in the equal mass case $\mu=1$ (bottom and top panels of Figure~\ref{fig:Ir1}, respectively).
We can also see that ${\cal I}$ grows monotonically with the interaction strength $g$.
The growth is what one would expect, since the invariant ${\cal I}$, as seen in the definition (\ref{eq:Ir1}), is a quantity that measures a width of the wave function in the relative coordinate (the width grows with increasing repulsions).
There is one exception for the totally antisymmetric states (so-called Girardeau), which do not depend on the interaction $g$.
They exist only in the equal mass ($\mu=1$) mixtures and are visible in the spectra of Figure~\ref{fig:spectrumM1} as horizontal lines.
In these states, due to the antisymmetry of the wave function under exchange of any two particles, the interaction energy vanishes.

Finally, let us note that it is possible also to define other invariants calculated from the many-body correlations.
For example, instead of ${\cal I}$, one can consider an existence of a different quantity ${\cal J}$ related to the width of the centre of mass distribution of a state $| i \rangle$:
\begin{equation}
  {\cal J} = \langle i | \hat{R}_{\mathrm{CM}}^2 | i \rangle.
\end{equation}
This quantity, in contrast to ${\cal I}$, is invariant under the change of excitations of the relative coordinates, i.e. it does not depend on the quantum numbers ($\nu_2,\nu_3,\ldots,\nu_N$).
Therefore, it can be used as an indicator of excitation of the centre of mass regardless of the internal motion.
It can be shown that, in contrast to ${\cal I}$, which depends only on the two-body pair-density profile, ${\cal J}$ depends also on single-particle densities and pair-density profiles of the fermions of the same type:
\begin{equation}
{\cal J} = \sum_{\sigma} N_\sigma \mu_\sigma^2\int \rho_\sigma(x)x^2 \mathrm{d}x
 + \sum_{\sigma} N_\sigma (N_\sigma - 1) \int C_{\sigma \sigma}(x,x') x x' \mathrm{d}x \mathrm{d}x'
 + \mu N_\downarrow N_\uparrow \int C (x,y) \mathrm{d}x \mathrm{d}y,
\end{equation}
where the pair-density profile for the same type fermions is defined analogously to (\ref{eq:correlation}):
\begin{equation}
 C_{\sigma \sigma}(x,x') = \langle i| \hat{\Psi}_\sigma^\dag(x) \hat{\Psi}_\sigma^\dag(x') \hat{\Psi}_\sigma(x') \hat{\Psi}_\sigma(x) |i \rangle.
\end{equation}
Since the invariant ${\cal J}$ depends on all three two-body density profiles, it is a less convenient quantity than ${\cal I}$ for a practical use.

\section{Conclusions}\label{sec:conclusions}
We have studied the properties of the system of a mass imbalanced two-component fermionic mixture in the centre-of-mass frame.
We show how to numerically construct the Hamiltonian of the centre of mass $\hat{H}_{\mathrm{CM}}$ and how to obtain eigenvectors of both Hamiltonians $\hat{H}$ and $\hat{H}_{\mathrm{CM}}$ expressed in a natural Fock basis.
Then, by determining excitations of the centre-of-mass motion, we filtered out its excitations and obtained simplified spectrum of the system being in the ground state of the centre-of-mass motion.
In this way we were able to perform analysis of many different properties of the system, regardless of the excitations of the centre of mass.
The procedure has been applied to systems with different number of fermions and different mass ratio generalizing previous results \cite{Busch1998,loft2015variational}.

To investigate the properties of a system confined in a harmonic trap we have shown that eigenstates are characterized by specific invariants that can be straightforwardly calculated from the observable available in the experiment.
The fact that some important information about the eigenstates can be extracted from the experimentally accessible correlations comes straightforwardly from from the separation of the centre-of-mass motion.

By comparing the values of the invariant ${\cal I}$ calculated theoretically with these measured experimentally one can control and engineer output eigenstates in experiments.
Moreover, both methods (filtering out the centre of mass excitations and using invariants encoded in pair correlations) can be very useful in distinguishing eigenstates labeled by different quantum numbers ${\bm \nu}$.
Therefore, the method can be used for analysing experimental outputs as long as the eigenstates of the system are consdered.
If the output state is a superposition or a mixed state, then the method is still valid, but might be not as helpful as in the case of eigenstates.

Let us note that the results presented may be important also in the case of deviations from a harmonic potential.
Then the centre-of-mass motion is coupled to the relative motion of the particles and can lead to very interesting phenomena known as Inelastic Confinement-Induced Resonances that has been recently studied both theoretically and experimentally \cite{Sowinski2012Creation,Saenz2012ICIR,Saenz2013ICIR}.

We believe that our results may shed some light on the strongly correlated few-body systems regardless the centre of mass excitations.
In broader context, the above methods may be used in other fields, like nuclear physics  \cite{Ozen2014mappingNuclear,Zinner2013nucleiColdAtoms} or chemistry.
The giant numerical complexity of the many-body systems increases exponentially with the number of particles, therefore the limit for our scheme is set by the numerical power of computers.
The method can be easily generalized to Bose-Bose and Fermi-Bose mixtures and can be used beyond the short-range  contact potential as long as the form of the interatomic potential allows one to separate out the centre-of-mass motion.

\begin{acknowledgements}
This work was supported by the (Polish) National Science Center Grants No. 2016/21/N/ST2/03315 (DP) and 2016/22/E/ST2/00555 (TS). MG acknowledges support from the EU Horizon 2020-FET QUIC641122.
\end{acknowledgements}

\bibliographystyle{spphys}       
\bibliography{references}   

\begin{thebibliography}{10}
\providecommand{\url}[1]{{#1}}
\providecommand{\urlprefix}{URL }
\expandafter\ifx\csname urlstyle\endcsname\relax
  \providecommand{\doi}[1]{DOI \discretionary{}{}{}#1}\else
  \providecommand{\doi}{DOI \discretionary{}{}{}\begingroup
  \urlstyle{rm}\Url}\fi

\bibitem{Blume2010TwoThreeToMany}
D.~Blume, Physics \textbf{3}, 74 (2010)

\bibitem{Blume2012FewBodyTraps}
D.~Blume, Rep. Prog. Phys. \textbf{75}, 046401 (2012)

\bibitem{VolosnievNC2014}
A.G. {Volosniev}, D.V. {Fedorov}, A.S. {Jensen}, M.~{Valiente}, N.T. {Zinner},
  Nature Communications \textbf{5}, 5300 (2014).

\bibitem{lewenstein2012ultracold}
M.~Lewenstein, A.~Sanpera, V.~Ahufinger, \emph{{Ultracold Atoms in Optical
  Lattices: Simulating quantum many-body systems}} (Oxford University Press,
  Oxford, 2012)

\bibitem{wenz2013fewToMany}
A.N. Wenz, G.~Z{\"u}rn, S.~Murmann, I.~Brouzos, T.~Lompe, S.~Jochim, Science
  \textbf{342}(6157), 457 (2013).

\bibitem{zurn2012fermionization}
G.~{Z{\"u}rn}, F.~{Serwane}, T.~{Lompe}, A.N. {Wenz}, M.G. {Ries}, J.E. {Bohn},
  S.~{Jochim}, Physical Review Letters \textbf{108}, 075303 (2012).

\bibitem{serwane2011tunableFermionSystem}
F.~Serwane, G.~Z{\"u}rn, T.~Lompe, T.B. Ottenstein, A.N. Wenz, S.~Jochim,
  Science \textbf{332}, 336 (2011).

\bibitem{MurmannPRL2015a}
S.~Murmann, F.~Deuretzbacher, G.~Z\"urn, J.~Bjerlin, S.M. Reimann, L.~Santos,
  T.~Lompe, S.~Jochim, Phys. Rev. Lett. \textbf{115}, 215301 (2015).

\bibitem{Kaufman2015Entangling}
A.M. Kaufman, B.J. Lester, M.~{Foss-Feig}, M.L. Wall, A.M. Rey, C.~Regal,
  Nature \textbf{527}, 208 (2015)

\bibitem{Onofrio2016Mixtures}
R.~Onofrio, Physics-Uspekhi \textbf{59}, 1129 (2017).

\bibitem{Busch1998}
T.~Busch, B.G. Englert, K.~Rz\c{a}{\.z}ewski, M.~Wilkens, Found. Phys.
  \textbf{28}, 549 (1998)

\bibitem{LiuHuDrummond2010ThreeAttractive}
X.J. Liu, H.~Hu, P.D. Drummond, Phys. Rev. A \textbf{82}, 023619 (2010)

\bibitem{KestnerDuan2007ThreeBody}
J.P. Kestner, L.M. Duan, Phys. Rev. A \textbf{76}, 033611 (2007)

\bibitem{WernerCastin2006UnitaryThreeBody}
F.~Werner, Y.~Castin, Phys. Rev. Lett. \textbf{97}, 150401 (2006)

\bibitem{loft2015variational}
N.J.S. Loft, A.S. Dehkharghani, N.P. Mehta, A.G. Volosniev, N.T. Zinner, EPJ D
  \textbf{69}, 65 (2015)

\bibitem{Koscik2012}
P.~Ko{\'s}cik, Eur. Phys. J. B \textbf{85}, 173 (2012)

\bibitem{Koscik2015vonNeuman}
P.~Ko{\'s}cik, Phys. Lett. A \textbf{379}, 293 (2015)

\bibitem{Koscik2016Bipartite}
P.~Ko{\'s}cik, Phys. Lett. A \textbf{380}, 1256 (2016)

\bibitem{Olshanii20154Body}
M.~Olshanii, S.G. Jackson, New J. Phys. \textbf{17}, 105005 (2015)

\bibitem{Harshman2016I}
N.L. Harshman, Few-Body Syst. \textbf{57}, 11 (2016)

\bibitem{Harshman2016II}
N.L. Harshman, Few-Body Syst. \textbf{57}, 45 (2016)

\bibitem{Dehkharghani2016Impenetrable}
A.S. Dehkharghani, A.G. Volosniev, N.T. Zinner, Journal of Physics B: Atomic,
  Molecular and Optical Physics \textbf{49}, 085301 (2016).

\bibitem{Garcia-MarchPRA2014}
M.A. Garc\'{\i}a-March, B.~Juli\'a-D\'{\i}az, G.E. Astrakharchik, J.~Boronat,
  A.~Polls, Phys. Rev. A \textbf{90}, 063605 (2014).

\bibitem{Gajda2006}
M.~Gajda, Phys. Rev. A \textbf{73}, 023603 (2006)

\bibitem{lawson1974}
D.H. Gloeckner, R.D. Lawson, Phys. Lett. B \textbf{53}, 313 (1974)

\bibitem{FedorovMikkelsen2015Recombination}
D.V. Fedorov, M.~Mikkelsen, A.S. Jensen, N.T. Zinner, Few-Body Syst.
  \textbf{56}, 889 (2015)

\bibitem{Amin2016CorrelationsHO}
R.E. Barfknecht, A.S. Dehkharghani, A.~Foerster, N.T. Zinner, J. Phys. B: At.
  Mol. Opt. Phys. \textbf{49}, 135301 (2016)

\bibitem{Gajda2000COM}
M.A. Za\l{}uska-Kotur, M.~Gajda, A.~Or\l{}owski, J.~Mostowski, Phys. Rev. A
  \textbf{61}, 033613 (2000)

\bibitem{Bienias2014Rydberg}
P.~Bienias, S.~Choi, O.~Firstenberg, M.F. Maghrebi, M.~Gullans, M.D. Lukin,
  A.V. Gorshkov, H.P. B\"uchler, Phys. Rev. A \textbf{90}, 053804 (2014)

\bibitem{Brandt2016COMconvergence}
J.G. Cosme, C.~Weiss, J.~Brand, Phys. Rev. A \textbf{94}, 043603 (2016)

\bibitem{Blakie2016PCS}
L.A. Williamson, P.B. Blakie, Phys. Rev. A \textbf{94}, 063615 (2016)

\bibitem{Bradley2016Brownian}
R.G. McDonald, A.S. Bradley, Phys. Rev. A \textbf{93}, 063604 (2016)

\bibitem{Wille6Li40K}
E.~Wille, F.~Spiegelhalder, G.~Kerner, D.~Naik, A.~Trenkwalder, G.~Hendl,
  F.~Schreck, R.~Grimm, T.G. Tiecke, J.T.M. Walraven, S.J.J.M.F. Kokkelmans,
  E.~Tiesinga, P.S. Julienne, Phys. Rev. Lett. \textbf{100}, 053201 (2008)

\bibitem{tiecke2010Feshbach6Li40K}
T.G. Tiecke, M.R. Goosen, A.~Ludewig, S.D. Gensemer, S.~Kraft, S.J.J.M.F.
  Kokkelmans, J.T.M. Walraven, Phys. Rev. Lett. \textbf{104}, 053202 (2010)

\bibitem{PethickSmith}
C.J. Pethick, H.~Smith, \emph{{Bose-Einstein condensation in dilute gases}}
  (Cambridge University Press, Cambridge, 2008)

\bibitem{Olshanii1998}
M.~Olshanii, Phys. Rev. Lett. \textbf{81}, 938 (1998).

\bibitem{IBB1985COM}
I.~Bialynicki-Birula, Lett. Math. Phys. \textbf{10}, 189 (1985)

\bibitem{IBB2002COM}
I.~Bialynicki-Birula, Z.~Bialynicka-Birula, Phys. Rev. A \textbf{65}, 063606
  (2002)

\bibitem{Sowinski2007Linear}
T.~Sowi{\'n}ski, Acta Phys. Polon. \textbf{38}, 2173 (2007)

\bibitem{SowinskiGrass2013FewInteracting}
T.~Sowi{\'n}ski, T.~Grass, O.~Dutta, M.~Lewenstein, Phys. Rev. A \textbf{88},
  033607 (2013).

\bibitem{Gharashi2013UpperBranchCorrelations}
S.E. Gharashi, D.~Blume, Phys. Rev. Lett. \textbf{111}, 045302 (2013).

\bibitem{Pecak2016Separation}
D.~P{\k e}cak, M.~Gajda, T.~Sowi{\'n}ski, New Journal of Physics \textbf{18},
  013030 (2016).

\bibitem{Pecak2016Transition}
D.~P{\k e}cak, T.~Sowi{\'n}ski, Phys. Rev. A \textbf{94}, 042118 (2016)

\bibitem{Manz2010G2}
S.~Manz, R.~B\"ucker, T.~Betz, C.~Koller, S.~Hofferberth, I.E. Mazets,
  A.~Imambekov, E.~Demler, A.~Perrin, J.~Schmiedmayer, T.~Schumm, Phys. Rev. A
  \textbf{81}, 031610 (2010).

\bibitem{bakr2010probing}
W.S. Bakr, A.~Peng, M.E. Tai, R.~Ma, J.~Simon, J.I. Gillen, S.~Foelling,
  L.~Pollet, M.~Greiner, Science \textbf{329}, 547 (2010)

\bibitem{sherson2010single}
J.F. Sherson, C.~Weitenberg, M.~Endres, M.~Cheneau, I.~Bloch, S.~Kuhr, Nature
  \textbf{467}, 68 (2010)

\bibitem{OmranBloch2015PauliBlocking}
A.~Omran, M.~Boll, T.A. Hilker, K.~Kleinlein, G.~Salomon, I.~Bloch, C.~Gross,
  Phys. Rev. Lett. \textbf{115}, 263001 (2015)

\bibitem{Cheuk2015FermionicMicroscope}
L.W. Cheuk, M.A. Nichols, M.~Okan, T.~Gersdorf, V.V. Ramasesh, W.S. Bakr,
  T.~Lompe, M.W. Zwierlein, Phys. Rev. Lett. \textbf{114}, 193001 (2015)

\bibitem{Cheuk2016MottMicroscope}
L.W. Cheuk, M.A. Nichols, K.R. Lawrence, M.~Okan, H.~Zhang, M.W. Zwierlein,
  Phys. Rev. Lett. \textbf{116}, 235301 (2016)

\bibitem{Cheuk2016Correlations}
L.W. Cheuk, M.A. Nichols, K.R. Lawrence, M.~Okan, H.~Zhang, E.~Khatami,
  N.~Trivedi, T.~Paiva, M.~Rigol, M.W. Zwierlein, Science \textbf{353}, 1260
  (2016)

\bibitem{ParsonsPRL2015}
M.F. Parsons, F.~Huber, A.~Mazurenko, C.S. Chiu, W.~Setiawan, K.~Wooley-Brown,
  S.~Blatt, M.~Greiner, Phys. Rev. Lett. \textbf{114}, 213002 (2015).

\bibitem{EdgePRA2015ImagingFermions}
G.J.A. Edge, R.~Anderson, D.~Jervis, D.C. McKay, R.~Day, S.~Trotzky, J.H.
  Thywissen, Phys. Rev. A \textbf{92}, 063406 (2015).

\bibitem{haller2015single}
E.~Haller, J.~Hudson, A.~Kelly, D.A. Cotta, B.~Peaudecerf, G.D. Bruce, S.~Kuhr,
  Nature Physics \textbf{11}, 738 (2015)

\bibitem{Gajda2016Imaging}
M.~Gajda, J.~Mostowski, T.~Sowi\'nski, M.~Za\l{}uska-Kotur, Europhys. Lett.
  \textbf{115}, 20012 (2016)

\bibitem{Sowinski2012Creation}
T.~Sowi\'{n}ski, Phys. Rev. Lett. \textbf{108}, 165301 (2012)

\bibitem{Saenz2012ICIR}
S.~Sala, P.I. Schneider, A.~Saenz, Phys. Rev. Lett. \textbf{109}, 073201 (2012)

\bibitem{Saenz2013ICIR}
S.~Sala, G.~Z\"urn, T.~Lompe, A.N. Wenz, S.~Murmann, F.~Serwane, S.~Jochim,
  A.~Saenz, Phys. Rev. Lett. \textbf{110}, 203202 (2013)

\bibitem{Ozen2014mappingNuclear}
C.~{\"O}zen, N.T. Zinner, EPJ D \textbf{68}, 225 (2014)

\bibitem{Zinner2013nucleiColdAtoms}
N.T. Zinner, A.S. Jensen, Journal of Physics G: Nuclear and Particle Physics
  \textbf{40}, 053101 (2013).

\end{thebibliography}

\end{document}